\documentclass[aps,prb,twocolumn,superscriptaddress,a4paper,nofootinbib]{revtex4-2}
\usepackage[dvipdfmx]{graphicx}

\usepackage{amsmath,amsthm,amssymb}
\usepackage{amsmath,bm}
\usepackage{color}
\usepackage{ifthen}

\usepackage{multirow,bigdelim}

\newcount \commentout
\newcommand{\1}{\mbox{1}\hspace{-0.25em}\mbox{l}}

\newlength{\figwidth}
\newlength{\figlarge}
\begin{document}
\title{
Chiral edge modes in evolutionary game theory: a kagome network of rock-paper-scissors
}

\author{Tsuneya Yoshida}
\email{yoshida@rhodia.ph.tsukuba.ac.jp}
\author{Tomonari Mizoguchi}
\author{Yasuhiro Hatsugai}
\affiliation{Department of Physics, University of Tsukuba, Ibaraki 305-8571, Japan}
\date{\today}
\begin{abstract}
We theoretically demonstrate the realization of a chiral edge mode in a system beyond natural science.
Specifically, we elucidate that a kagome network of rock-paper-scissors (K-RPS) hosts a chiral edge mode of the population density which is protected by the non-trivial topology in the bulk.
The emergence of the chiral edge mode is demonstrated by numerically solving the Lotka-Volterra (LV) equation. 
This numerical result can be intuitively understood in terms of cyclic motion of a single rock-paper-scissors cycle which is analogous to the cyclotron motion of fermions. 
Furthermore, we point out that a linearized LV equation is mathematically equivalent to the Schr\"odinger equation describing quantum systems. 
This equivalence allows us to clarify the topological origin of the chiral edge mode in the K-RPS; a non-zero Chern number of the payoff matrix induces the chiral edge mode of the population density, which exemplifies the bulk-edge correspondence in two-dimensional systems described by evolutionary game theory.
\end{abstract}
\maketitle

\section{
Introduction
}
\label{sec: intro}
Notion of topology plays a central role in condensed matter physics~\cite{Thouless_PRL1982,Kane_Z2TI_PRL05_1,Kane_Z2TI_PRL05_2,HgTe_Bernevig06,Qi_TQFTofTI_PRB08,TI_review_Hasan10,TI_review_Qi10}. 
One of the remarkable properties of topological system is the emergence of edge states~\cite{Hatsugai_PRL93} protected by the topology in the bulk which is a source of anomalous behaviors.
For instance, integer quantum Hall systems show the quantized Hall conductance with extremely high accuracy~\cite{Ando_IQHE_JPSJ75,Klitzing_IQHE_PRL80,Thouless_PRL1982,Halperin_PRB82} due to the chiral edge mode (i.e., one-way propagating modes localized around the edge).

So far, insulators and superconductors have been extensively analyzed as platforms of topological physics. 
However, recently, it turned out that topological phenomena~\cite{Szolnoki_GameRev_JRSI14,Wang_GemeRevEPJB15}~\footnote{
References~\onlinecite{Szolnoki_GameRev_JRSI14,Wang_GemeRevEPJB15} mentions topology of game theory. We note, however, that topology discussed in these references differs from the topology of eigenvectors (, or eigenstates). 
One can characterize the topology of the eigenstates by topological invariant which is computed from eigenvectors.
}
extend beyond the quantum systems~\cite{Albert_Topoelecircit_PRL15,Ningyuan_Topoelecircit_PRX15,Victor_Topoelecircit_PRL15,Lee_Topoelecircit_CommPhys18,Ezawa_HOTIele_PRB18,Imhof_HOTI_NatPhys18,Yoshida_MSkinPRR20,Perrot_fluid_NatPhys19,Tauber_BBCcont_PRR20,Yoshida_topodiff_SciRep20,Knebel_RPSchain_PRL20}.
In particular, the chiral edge modes protected by topological properties have been reported for various systems, such as photonic crystals~\cite{Haldane_chiralPHC_PRL08,Raghu_chiralPHC_PRA08,Wang_chiralPHC_Nature09,Ozawa_TopoPhoto_RMP19,He_HOTI_photonic_NatCom20}, mechanical metamaterials~\cite{ProdanPRL09,Kane_NatPhys13,Kariyado_SR15,Susstrunk_TopoMech_Sci15,Nash_Mechchiral_PNAS15,Huber_TopoMech_NatPhys16,Suesstrunk_Mech-class_PNAS16,Serra-Garcia_HOTIphonon_Nat_2018,Noh_HOTIPhonon_NatPho2018,Ni_HOTIacostic_NatMat2019,Yoshida_SPERs_mech19,Xue_HOTIphonon_2019,Wakao_HOTImech_PRB20,Scheibner_nHSkinmech_PRL20,Wakao_Topoband_JPSJ2020,Xie_HOTIclassical_arXiv2020}, equatorial waves~\cite{Delplace_topoEq_Science17}, active matter~\cite{ActiveMatter_SonePRL19,Sone_ActiveMatter_NatComm20,Yamauchi_ActiveMatter_arXiv20}, and so on.
This progress is significant not only from an academic viewpoint but also from an engineering viewpoint because such topologically protected chiral edge modes may result in new inventions, e.g., the topological insulator laser~\cite{Harari_TopoLaser_Science18,Banders_TopoLaser_Science18} and potential application to a novel energy transfer system of extremely low transmission loss.

In spite of the above progress, the chiral edge modes are still restricted to systems of natural science. 
Discovery of chiral edge modes beyond the natural science is crucial as it may provide a new perspective.

\begin{figure}[!h]
\begin{minipage}{0.7\hsize}
\begin{center}
\includegraphics[width=1\hsize,clip]{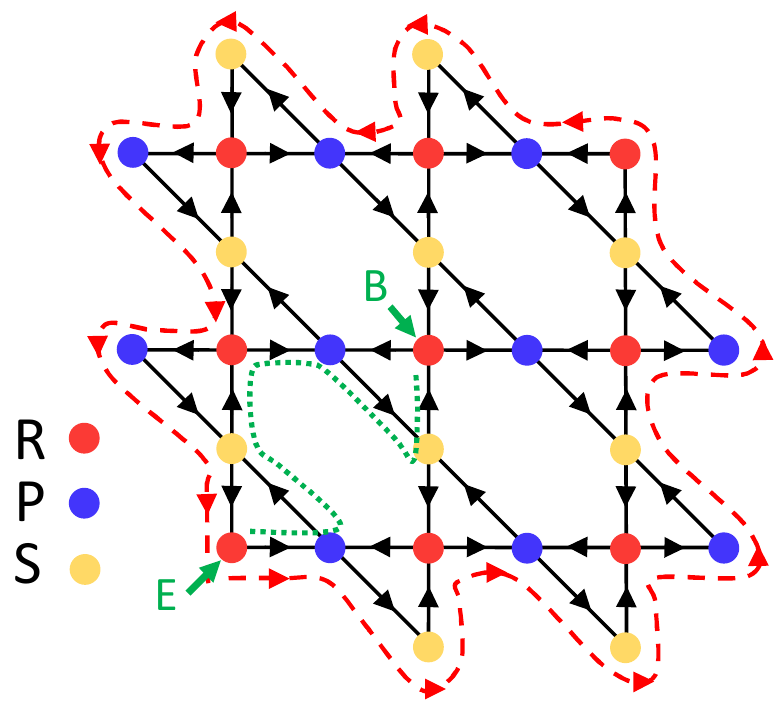}
\end{center}
\end{minipage}
\caption{
(Color Online).
Kagome network of rock-paper-scissors. 
At site $I$ strategy $s_I$ is assigned as illustrated by a colored dot; red, blue, and yellow dots denote strategies, rock (R), paper (P), scissors (S), respectively.
In this panel, the arrows connecting sites illustrate the payoffs; R beats S; S beats P; P beats R.
(The explicit form of the payoffs is shown in Eq.~(\ref{eq: A single RPS}) for a single RPS cycle.)
This kagome network of rock-paper-scissors is described by evolutionary game theory rather than by solid state physics. 
However, it can be mapped to a spinless fermion model [see Fig.~\ref{fig: K-ele}].
}
\label{fig: K-RPS}
\end{figure}

The aim of this paper is to report the discovery of a chiral edge mode in a system of evolutionary game theory which is beyond natural science~\cite{Rosas_Social_JTB10,Wang_SocSci_PRE13,Perc_Game_PhysRep17}~\footnote{
Evolutionary game theory can be applied not only to natural science but also to social science~\cite{Rosas_Social_JTB10,Wang_SocSci_PRE13,Perc_Game_PhysRep17}.
}.
Specifically, we elucidate that a chiral edge mode of the population density~\footnote{
For definition of the population density, see just below of Eq.~(\ref{eq: LV eq}).
} 
emerges in a kagome network of rock-paper-scissors (K-RPS)~\footnote{
Namely, players choose the given strategy assigned at each site as Fig.~\ref{fig: K-RPS}.
} 
[see Fig.~\ref{fig: K-RPS}] due to the non-trivial topology in the bulk.
On each site of the K-RPS, one of the strategies, rock (R), paper (P), and scissors (S), is assigned. 
The payoff of a player on a given site is described by the arrows of bonds connecting the sites.
The emergence of the chiral edge mode is demonstrated by the numerically solving the Lotka-Volterra (LV) equation.
This result can be intuitively understood by focusing on cyclic motion of the single rock-paper-scissors (RPS) cycle which is analogous to the cyclotron motion of fermions under a uniform magnetic field.
Furthermore, we elucidate the topological origin of the chiral edge modes by pointing out the mathematical equivalence of a linearized LV equation of the K-RPS and the Schr\"odinger equation of a fermionic kagome lattice model with the non-trivial topology [see Fig.~\ref{fig: K-ele}]. 
This equivalence elucidates the bulk-edge correspondence for two-dimensional systems described by evolutionary game theory; a non-zero Chern number of the payoff matrix induces the chiral edge modes observed in the time-evolution of the population density of players located at the sites of the K-RPS.

Topological band structure and a zero mode have been discussed for a RPS chain~\cite{Knebel_RPSchain_PRL20}. 
We would like to stress, however, that the presence of the chiral edge modes remains unsolved because Ref.~\onlinecite{Knebel_RPSchain_PRL20} analyzes the one-dimensional system. 
Novelty of our work is discovering the topological origin of the one-way mode by pointing out the mathematical equivalence of the linearized LV equation and the Sch\"odinger equation.

\section{
Cyclotron motion in a single RPS cycle
}
\label{sec: single RPS}
As a first step, we show that ``cyclotron motion" can be observed in a single RPS cycle~\cite{SZABO200797}, which plays a key role in the emergence of chiral edge modes beyond natural science.

\begin{figure}[!h]
\begin{minipage}{1\hsize}
\begin{center}
\includegraphics[width=0.8\hsize,clip]{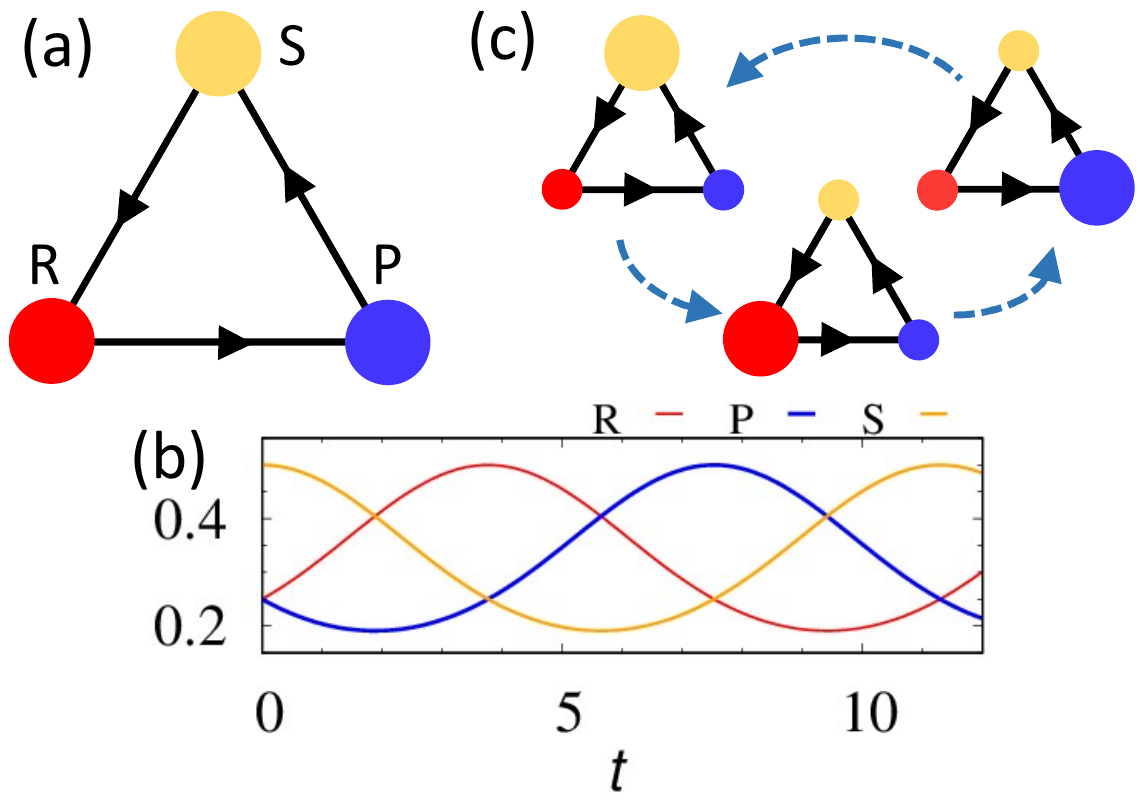}
\end{center}
\end{minipage}
\caption{
(Color Online). 
(a): Sketch of a RPS cycle. 
The arrows in panel (a) denote the dominance relationship between the strategies; R beats S; S beats P; P beats R.
(b): Dynamics of the RPS cycle with an initial condition $\bm{x}_{\mathrm{ini}}=(\frac{1}{4},\frac{1}{4},\frac{1}{2})$.
(c): Sketch of the dynamics of the RPS cycle. The size of the circle indicates the population density.
}
\label{fig: S-RPS}
\end{figure}

Consider two players who choose one of the strategies $(s_1,s_2,s_3) = (\mathrm{R},\mathrm{P},\mathrm{S})$. The rule of the game is illustrated in Fig.~\ref{fig: S-RPS}(a); R beats S; S beats P; P beats R.
In this case, the payoff of a player is $A_{IJ}$ if the player chooses the strategy $s_I$ and the other player choosing $s_J$ ($I,J=1,2,3$). Here the payoff matrix of this game is given by
\begin{eqnarray}
\label{eq: A single RPS}
A &=& 
\left(
\begin{array}{ccc}
0   &  -1  & 1 \\
1   &  0   & -1  \\
-1  &  1   & 0
\end{array}
\right).
\end{eqnarray}

Now, let us consider the case where a large number of players repeat the game.
It is known that the dynamics of this game is described by the LV equation~\cite{Sigmund_Game_AMS11,Knebel_RPSchain_PRL20}~\footnote{
If the players know the prediction and try to behave in a different way, the population density may not follow this equation. However, so far, 
it is assumed that Eq.~(\ref{eq: LV eq}) holds.
Namely, it has been assumed that the increase/decrease of the population density corresponds to the payoff~\cite{Sigmund_Game_AMS11}
}
\begin{eqnarray}
\label{eq: LV eq}
\partial_t x_I &=& x_I \bm{e}^T_I A\bm{x},
\end{eqnarray}
with $\bm{x}=(x_1,x_2,x_3)^T$ and $x_1$, $x_2$, and $x_3$ being the population density~\footnote{
As mentioned in the main text, we normalize the vector $\bm{x}$ so that $\sum_I x_I=1$ holds. In this sence, $x_I$ denotes the population density
} 
of players who choose the strategies R, P, and S, respectively.
We suppose that the vector $\bm{x}$ is normalized $\sum_{I} x_I=1$ unless otherwise noted.
The vector $\bm{e}_I$ is the unit vector whose $I$-th element takes one; $[\bm{e}_I]_J=\delta_{IJ}$.
When a player chooses the strategy $s_I$, the expectation value of the payoff is written as $\bm{e}^T_IA\bm{x}$.
Thus, Eq.~(\ref{eq: LV eq}) indicates that the population density $x_I$ increases in order to enhance the payoff.

We note that the vector $\bm{c}=(1,1,1)^T/3$ satisfies 
\begin{eqnarray}
\label{eq: Nash}
A\bm{c} &=& 0,
\end{eqnarray}
which means that $\bm{x}=\bm{c}$ is a stationary state. This vector describes a Nash equilibrium~\cite{Weibull_textbook97,Loertscher_NoESS_EL13}~\footnote{
It should be noted that the Nash equilibrium does not necessarily corresponds to evolutionary stable strategy. 
Indeed, it is known that there is no evolutionary stable strategy in the single-rock-paper-scissors cycle~\cite{Weibull_textbook97,Loertscher_NoESS_EL13}
}
 for a classical game theory where the strategy of players does not change over time and players play the game only once.
For the relation $A\bm{c}=0$, the following fact is essential: at each site, the number of bonds with the out-going arrow is equal to the number of bonds with the in-coming arrow. Because of this fact, for arbitrary $I$, the expectation value $\bm{e}^T_{I} A \bm{c}$ is zero, which results in the relation $A\bm{c}=0$. 
Noting these facts, we can also find the stationary state for the K-RPS.

A slight deviation from the stationary state $\bm{c}$ results in cyclic motion which is analogous to the cyclotron motion of fermions in the Landau levels.
The time-evolution with the initial state $\bm{x}_{\mathrm{ini}}=(\frac{1}{4},\frac{1}{4},\frac{1}{2})$ is shown in Fig.~\ref{fig: S-RPS}(b). Here, we choose $A_{21}=1$ as a unit of time [see Eq. (2)].
As illustrated in Fig.~\ref{fig: S-RPS}(c), the data shown in Fig.~\ref{fig: S-RPS}(b) indicate the cyclic motion, which can be intuitively understood by noticing that the population density propagates along the arrows illustrated in Fig.~\ref{fig: S-RPS}(a).
Namely, with the given initial state, $\bm{x}_{\mathrm{ini}}=(\frac{1}{4},\frac{1}{4},\frac{1}{2})$, players who choose strategy R gain the highest payoff (i.e., $\bm{e}^T_IA\bm{x}$ become maximum for $I=1$).
Thus, after time-evolution, the population density of players choosing strategy R increases. 
In a similar manner, we can see that the population density of players choosing strategy P increases in the next step.

The above results elucidate that the players of the single RPS cycle mimic the ``cyclotron motion" as if they were fermions in the Landau levels.
We note that analogy of cyclotron motion can be mathematically shown (see Appendix~\ref{sec: fermi app}). As we see in Sec.~\ref{sec: K-RPS}, the ``cyclotron motion" in the single RPS play an important role to search a system exhibiting a chiral edge mode.

\section{
Dynamics of the K-RPS
}
\label{sec: K-RPS}
A typical example of quantum systems exhibiting chiral edge modes is an integer quantum Hall system. Fermions in this two-dimensional system show the cyclotron motion which breaks time-reversal symmetry.
Keeping this fact in mind, one can expect the emergence of chiral edge modes (i.e., one-way propagating the population density localized around the edge) in a two-dimensional network constructed from the RPS cycles which mimic the cyclotron motion.
Specifically, we consider the K-RPS illustrated in Fig.~\ref{fig: K-RPS} which indeed hosts a chiral edge modes.
\subsection{
Numerical results
}
\label{sec: K-RPS num}

By solving the LV equation~(\ref{eq: LV eq}) numerically, we demonstrate the presence of chiral edge modes in the K-RPS. 
The payoff matrix $A\in M(N_{\mathrm{tot}},\mathbb{R})$ can be read off from the arrow assigned to each bond [see Fig.~\ref{fig: K-RPS}]. 
Here $N_{\mathrm{tot}}$ denotes the number of sites.
In order to analyze the time-evolution, we employ a fourth order Runge-Kutta method~\cite{Suli_RK_textbook03}.
We discretize time as $t_n=n \Delta t $ with $\Delta t=0.1$ and $n=0,1,2,\ldots$.

\begin{figure}[!h]
\begin{center}
\includegraphics[width=1\hsize]{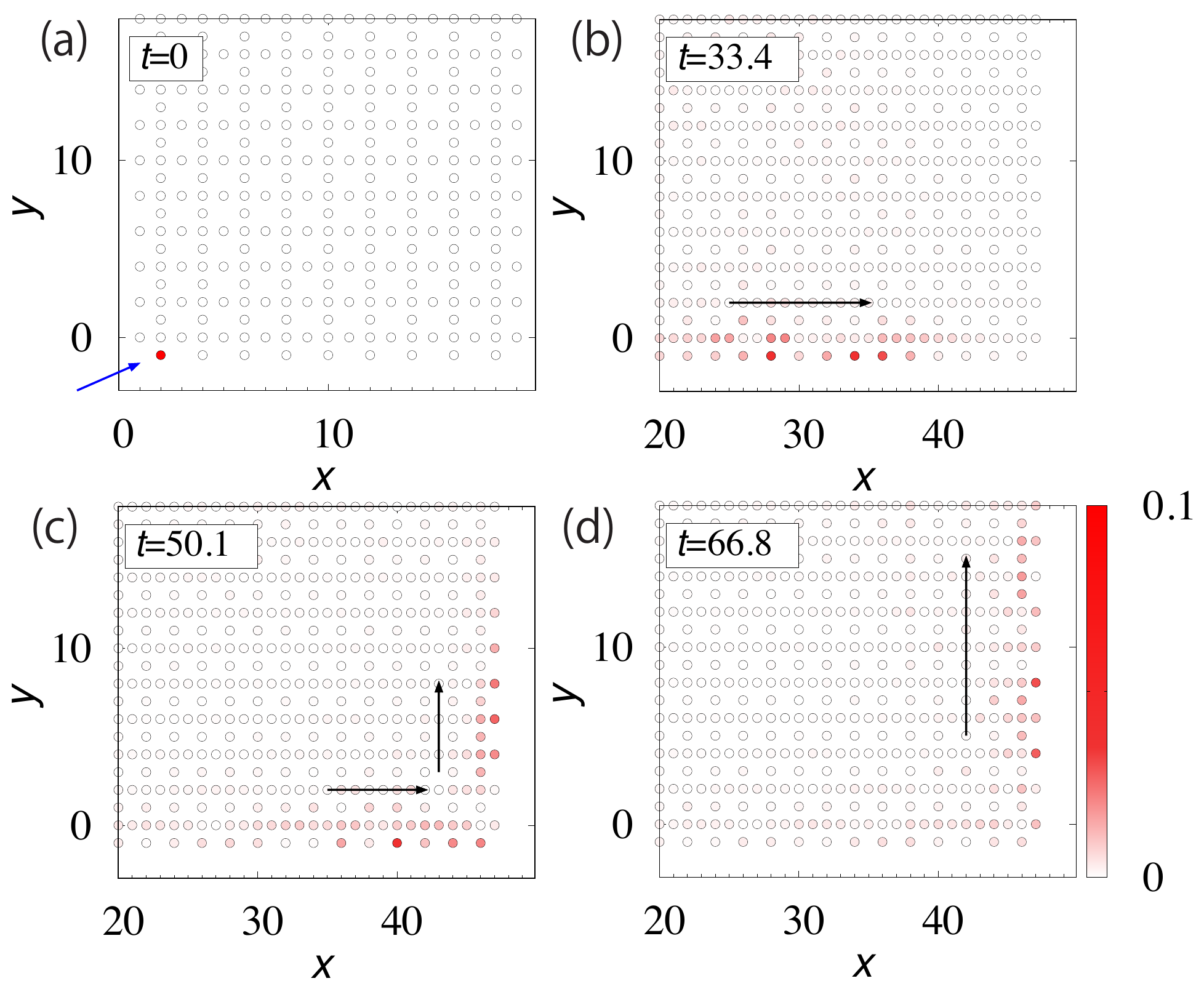}
\end{center}
\caption{
(Color Online).
Time-evolution of the population density for the K-RPS.
In these figures, the absolute value of the deviation $|\delta \bm{x}|$ ($\delta \bm{x}=\bm{x}-N_{\mathrm{tot}}\bm{c}^{(\mathrm{K})}$) is plotted; the vector $\bm{x}$ satisfies $\sum_I \bm{x}_I=N_{\mathrm{tot}}+0.1$.
For $t=0$, $\delta x_I$ takes 0.1 at the site denoted by the blue arrow in panel (a); otherwise $\delta x_I$ is zero.
The deviation of the population density propagates along the edge in the counter-clockwise direction as illustrated by the black arrows in panels (b)-(d).
We have simulated the time-evolution up to $t_{\mathrm{max}}=200$. For $0\leq t \leq t_{\mathrm{max}}$, the chiral edge mode is observed.
More detailed data are shown in Fig.~\ref{fig: Timeevol_K-RPS org app} of Appendix~\ref{sec: time evol org app}.
The scale of color plot in panels (a)-(c) is the same as the one in panel (d).
}
\label{fig: Timeevol_K-RPS}
\end{figure}

We analyze the dynamics with an initial state~\footnote{
We have also analyze evolution of an initial state which describes a population density deviating from $N_{\mathrm{tot}}\bm{c}^{(\mathrm{K})}$ only at the center of the system. In this case, we observe that the deviation propagates homogeneously
} 
which slightly deviates from a stationary state $\bm{c}^{(\mathrm{K})}=(1,1,1....,1)^T/N_{\mathrm{tot}}$ ($\bm{x}=\bm{c}^{(\mathrm{K})} +\delta \bm{x}$). 
The relation $A\bm{c}^{(\mathrm{K})}=0$ holds because the number of bonds with the in-coming arrow is equal to the number of bonds with the out-going arrow for each site of the K-RPS [see Fig.~\ref{fig: K-RPS} and the argument below Eq.~(\ref{eq: Nash})].

Figure~\ref{fig: Timeevol_K-RPS} clearly indicates the presence of the chiral edge mode; sites of the high population density (red dots) propagates along the edge in the counter-clockwise direction.
Our numerical results also imply topological stability of the chiral edge mode. In Appendix~\ref{sec: time evol org app}, we demonstrate the following two facts. (i) The system hosts the chiral modes even with an initial condition significantly deviating from $\bm{c}^{(\mathrm{K})}$ [see Fig.~\ref{fig: Timeevol_K-RPS app}]. (ii) Even in the presence of a defect on the edge, the chiral edge mode propagates by detouring around the defect [see Fig.~\ref{fig: Timeevol_K-RPS opp2 app}]. 
The latter result is particularly counter-intuitive.
The above results indicate the robustness of the chiral edge mode.

\subsection{
Intuitive discussion
}
\label{sec: K-RPS intuitive}
The dynamics obtained in Fig.~\ref{fig: Timeevol_K-RPS} is intuitively understood as follows.

Firstly, we recall that the K-RPS is composed of the single RPS cycle. 
This fact means that the population density propagates along the out-going arrows in order to maximize the payoff. 
For instance, when the population density at site denoted with ``E" is higher than the other sites [see Fig.~\ref{fig: K-RPS}], it propagates around the path illustrated in the red dashed line in Fig.~\ref{fig: K-RPS}, which implies the presence of the chiral edge modes.

The localization of the chiral edge mode can be deduced as follows. 
Firstly, we note that in the bulk, sites are connected by four bonds; out-going arrows are assigned to two of the bonds, and in-coming arrows are assigned to the other two bonds. Thus, in contrast to the players on the edge sites, those in the bulk have two options of out-going arrows which results in the localization of the chiral mode around the edge.
For instance, when the population density at site ``E" on the edge propagates to site ``B" in the bulk along the path denoted with green dashed-line in Fig.~\ref{fig: K-RPS}, it passes through five branches. Because the propagation is suppressed at each of these branches, the deviation of the population density is localized around the edges.

The above results explicitly demonstrates the emergence of the chiral edge mode in the K-RPS. 
As we see in Sec.~\ref{sec: topological K-RPS cheracterize}, the topology in the bulk governs the chiral edge modes.
This is supported by a simulation showin in Fig.~\ref{fig: Timeevol_K-RPS opp app} of Appendix~\ref{sec: time evol org app} where the payoff matrix is flipped, $A\to -A$, (Namely, we impose the opposite rule instead of the ordinary one; R beats P; P beats S; S beats R). 
The obtained data show the edge mode propagating the opposite direction.

\section{
Topological characterization of the chiral edge modes of the K-RPS
}
\label{sec: topological K-RPS}
So far, we have seen that the K-RPS hosts a chiral edge mode.
Here, we elucidate the topological origin by pointing out a relation between the K-RPS [Fig.~\ref{fig: K-RPS}] and the fermionic quantum model [Fig.~\ref{fig: K-ele}] with the non-trivial topology.

\subsection{
A linearized LV equation and the Schr\"odinger equation
}
\label{sec: topological K-RPS map}

In order to see the relation between the K-RPS and the fermionic kagome lattice model, we linearize the LV equation around the stationary state $\bm{c}^{(\mathrm{K})}$.

For $\bm{x}$ slightly deviating from $\bm{c}^{(\mathrm{K})}=(1,1,1,\ldots,1)^T/N_{\mathrm{tot}}$, the LV equation is rewritten as
\begin{eqnarray}
\partial_t \bm{e}_I \cdot \delta\bm{x} &=& \bm{e}_I \cdot (\bm{c}^{(\mathrm{K})}+\delta \bm{x}) \bm{e}^T_I A (\bm{c}+\delta \bm{x}) \nonumber \\
                                   &=&  (\bm{c}^{(\mathrm{K})}+\delta \bm{x}) \cdot \bm{e}_I \bm{e}^T_I A (\delta \bm{x}) \nonumber \\
                                   &\sim&  (\bm{c}^{(\mathrm{K})})^T P_I A \delta \bm{x}, 
\end{eqnarray}
with $P_I=\bm{e}_I\bm{e}^T_I$.
From the first to the second line, we have used the relation $A \bm{c}^{(\mathrm{K})} = 0$. In the last line, we have discarded the second order term of $(\delta \bm{x})^2$.

Noting the relation $N_{\mathrm{tot}} P_I \bm{c}^{(\mathrm{K})}=\bm{e}_I$, we can write the above linearized equation as
\begin{eqnarray}
\label{eq: d x = H x}
i\partial_t \delta\bm{x} &=&  \frac{1}{N_{\mathrm{tot}}} H \delta \bm{x},
\end{eqnarray}
with a Hermitian matrix $H= iA$, which is mathematically equivalent to the Schr\"odinger equation up to the prefactor.

\begin{figure}[!h]
\begin{minipage}{0.7\hsize}
\begin{center}
\includegraphics[width=1\hsize]{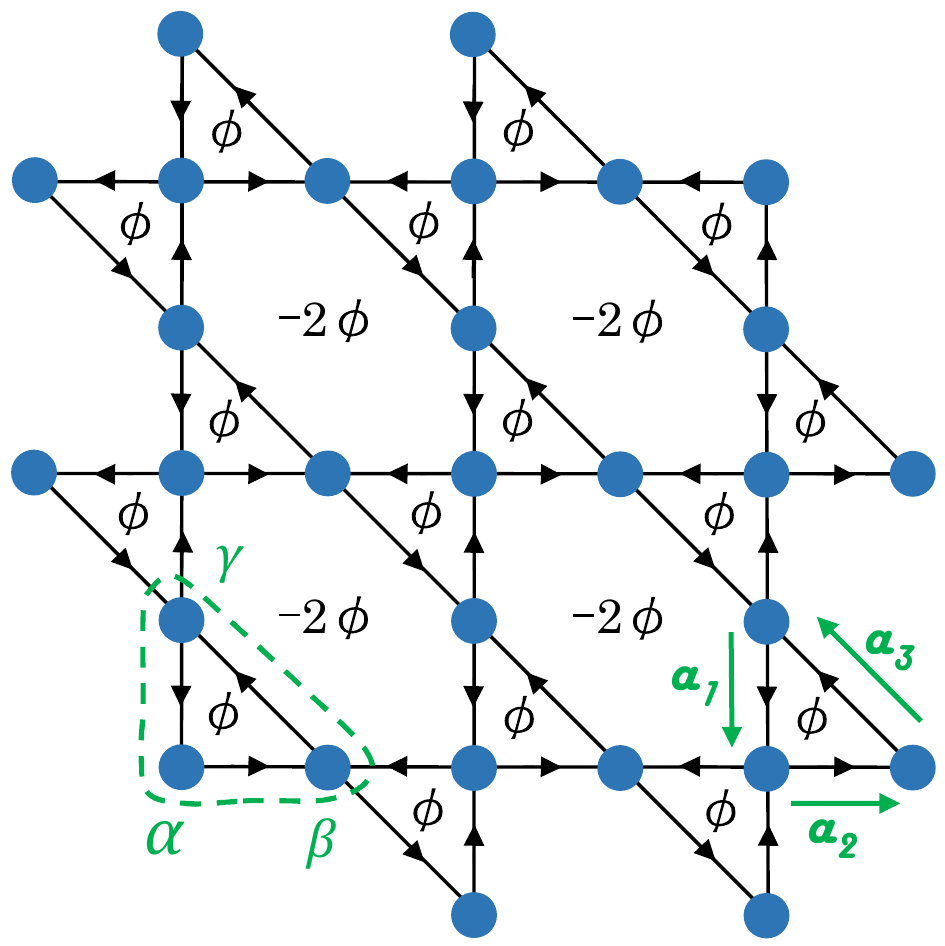}
\end{center}
\end{minipage}
\caption{
(Color Online).
Sketch of fermionic lattice model with magnetic fluxes introduced in Ref.~\onlinecite{Ohgushi_ChiralKagome_PRB00}. a spinless fermion acquires a phase $\phi$ when it hops around a triangle along the arrows.
The K-RPS shown in Fig.~\ref{fig: K-RPS} is mapped to the fermionic model with $\phi=3\pi/2$.
}
\label{fig: K-ele}
\end{figure}

Taking into account the explicit form of the payoff matrix $A$, we can see that the K-RPS is mapped to the tight-binding model of spinless fermion in the kagome lattice which hosts a chiral edge mode due to the non-trivial topology in the bulk.
To see this more clearly, let us discuss the following eigenvalue problem which governs the dynamics described by the linearized LV equation~(\ref{eq: d x = H x}):
\begin{subequations}
\label{eq: Bloch A}
\begin{eqnarray}
\sum_j A_{ij}(\bm{k})\psi_{jn} &=& \psi_{in}(\bm{k}) \epsilon_n,
\end{eqnarray}
\begin{eqnarray}
A(\bm{k})
&=&
\left(
\begin{array}{ccc}
 0                                 & -(1+e^{i2\bm{k} \cdot \bm{a}_2})  &   (1+e^{-2i\bm{k} \cdot \bm{a}_1})  \\
(1+e^{-i2\bm{k} \cdot \bm{a}_2})   & 0                                 & -(1+e^{2i\bm{k} \cdot \bm{a}_3})  \\
-(1+e^{2i\bm{k} \cdot \bm{a}_1})   & (1+e^{-2i\bm{k} \cdot \bm{a}_3})   & 0 
\end{array}
\right),\nonumber \\
\end{eqnarray}
\end{subequations}
where $A(\bm{k})$ with $\bm{k}=(k_x,k_y)$ is the Fourier transformed payoff matrix.
Here, $\psi_{jn}$ ($j,n=1,2,3$) denotes the $j$-th component of the eigenvector $\bm{\psi}_n$ with the eigenvalue $\epsilon_n$.
The vectors connecting the neighboring site are defined as $\bm{a}_{1}:=(0,-1)$, $\bm{a}_{2}:=(1,0)$, and $\bm{a}_{3}:=(-1,1)$ [see also Fig.~\ref{fig: K-ele}].
Because $A(\bm{k})$ is anti-Hermitian [$A(\bm{k})=-A^\dagger(\bm{k})$], the eigenvalues are pure imaginary ($\epsilon_n  \in i\mathbb{R} $).
We note that the Hermitian matrix $H(\bm{k})=iA(\bm{k})$ is identical to the Bloch Hamiltonian of spinless fermions in the kagome lattice with magnetic fluxes $\phi=3\pi/2$ (for details of the fermionic system see Appendix~\ref{sec: fermi app}).
The symmetry class of $H(\bm{k})$ is class D where particle-hole symmetry is preserved~\cite{Schnyder_classification_free_2008,Kitaev_classification_free_2009,Ryu_classification_free_2010}.

The above results clarify the mathematical equivalence of the linearized LV equation of the K-RPS and the Schr\"odinger equation of the fermionic kagome lattice model with the non-trivial topology~\footnote{
We stress that the above argument elucidates just the mathematical equivalence of the linearized LV equation and the Schr\"odinger equation. Thus, the equivalence does not mean that the population density itself obeys the Schr\"odinger equation
}.
This equivalence and the bulk-edge correspondence~\cite{Hatsugai_PRL93} provides a topological perspective of the chiral edge modes of the K-RPS whose details are discussed below.

\subsection{
Topological characterization of the chiral edge modes of the K-RPS
}
\label{sec: topological K-RPS cheracterize}
Based on the above results, we address the topological characterization of the chiral edge mode in the K-RPS.

Specifically, we characterize the edge mode by the following steps. 
Firstly, we analyze the bulk band structure as well as the Chern number [see Eq.~(\ref{eq: N_Ch})]. 
Secondly, by diagonalizing the system under the cylinder geometry, we demonstrate that the topological properties in the bulk induce the chiral edge mode, which is known as the bulk-edge correspondence~\cite{Hatsugai_PRL93} in the context of topological insulators/superconductors.
Here, the system under the cylinder geometry denotes the system where the periodic (open) boundary condition is imposed along the $x$- ($y$-) direction.

\begin{figure}[!h]
\begin{minipage}{1\hsize}
\begin{center}
\includegraphics[width=1\hsize,clip]{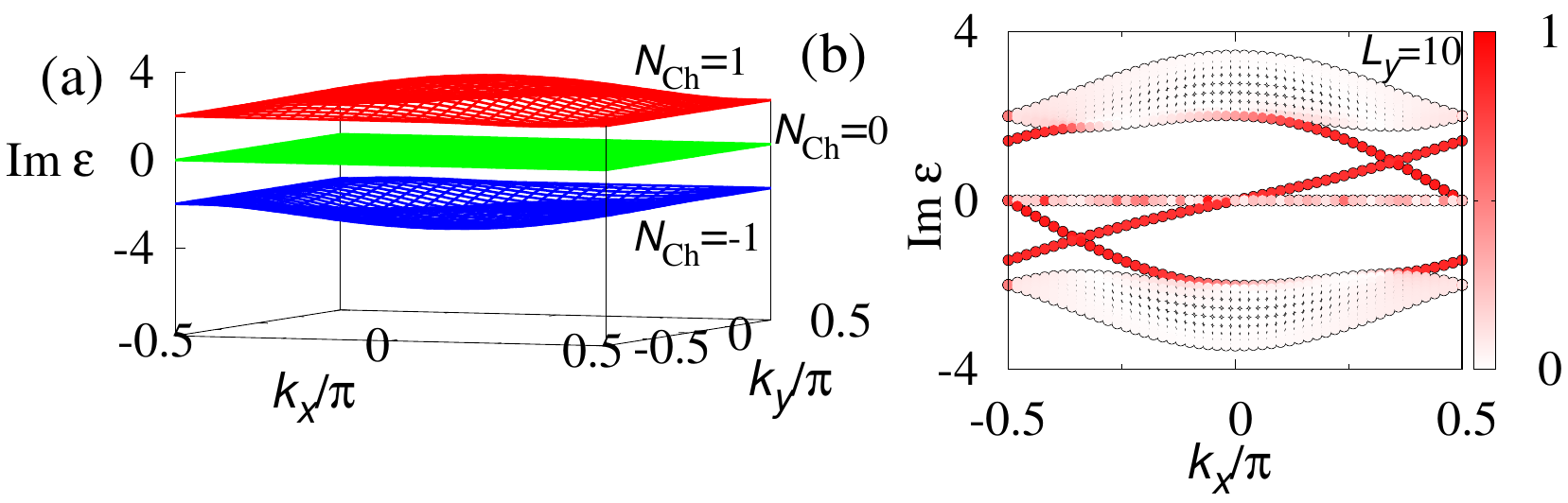}
\end{center}
\end{minipage}
\caption{
(Color Online).
(a): Band structure of ``Bloch Hamiltonian" $A(\bm{k})$. 
The Chern number takes $N_{\mathrm{Ch}}=-1$, $0$, and $1$ for the bottom, middle, and top bands, respectively.
(b): Spectrum of the K-RPS under the cylinder geometry. Color of data points represents the absolute value of the polarization $|P_n|$ defined in Eq.~(\ref{eq: P_n}).
Polarization for each eigenstate is plotted in Fig.~\ref{fig: band structures app} of Appendix~\ref{sec: Kagome_Pol app}.
}
\label{fig: band structures}
\end{figure}

Firstly, we discuss the bulk properties by diagonalizing the Fourier transformed payoff matrix $A(\bm{k})$ which is anti-Hermitian.
Figure~\ref{fig: band structures}(a) plots the band structure of $A(\bm{k})$.
Because each band is separated by a gap in the two-dimensional Brillouin zone, the Chern number of each band is quantized which is defined as
\begin{subequations}
\label{eq: N_Ch}
\begin{eqnarray}
N_{\mathrm{Ch}}&=& \sum_{\mu\nu} \epsilon_{\mu\nu} \int \! \frac{dk_xdk_y}{2\pi i} \partial_{k_\mu} \mathcal{A}_{n\nu},
\end{eqnarray}
\begin{eqnarray}
\mathcal{A}_{n\mu} &=& \sum_{j} \psi^\dagger_{nj}(\bm{k}) \partial_{k_\mu}  \psi_{jn}(\bm{k}),
\end{eqnarray}
\end{subequations}
Here, $\epsilon_{\mu\nu}$ ($\mu,\nu=x,y$) denotes the anti-symmetric tensor satisfying $\epsilon_{xy}=1$.

Employing the method based on Ref.~\onlinecite{Fukui_FHS_mechod_JPSJ05}, we find that the Chern number of the bottom (top) band takes $-1$ ($1$) [see Fig.~\ref{fig: band structures}(a)].

The non-trivial topology characterized by the Chern number in the bulk predicts the chiral edge modes around the boundary. 
Indeed, the spectrum of the matrix $A$ under the cylinder geometry exhibits the chiral edge modes [see Fig.~\ref{fig: band structures}(b)].
In Fig.~\ref{fig: band structures}(b) the color assigned to each eigenvalue $\epsilon_n$ denotes the absolute value of the polarization $|P_n|$ of the corresponding eigenvector $\bm{\psi}_{n}(k_x)$
\begin{eqnarray}
\label{eq: P_n}
P_n &=& 1- \frac{2}{L_y} \sum_{J_y}  \psi^\dagger_{nJ_y}(k_x) J_y \psi_{j_yn}(k_x),
\end{eqnarray}
where $J_y$ ($J_y=1,2,...,L_y$) labels the sites along the $y$-direction.
Figure~\ref{fig: band structures}(b) indicates that the chiral modes denoted by red-colored dots are localized around edges (for more details, see Fig.~\ref{fig: band structures app} of Appendix~\ref{sec: Kagome_Pol app}).
The replacement $A\to -A$ flips the sign of Chern number. 
Correspondingly, the direction of the chiral edge mode changes as mentioned in Sec.~\ref{sec: K-RPS num} (see Fig.~\ref{fig: Timeevol_K-RPS opp app} in Appendix~\ref{sec: time evol org app}). These behaviors correspond to the behaviors of the fermionic lattice model (Fig.~\ref{fig: K-ele}) where the magnetic fluxes are flipped ($\phi\to-\phi$).

The above results elucidate that the chiral edge mode of the population density in the K-RPS is induced by the non-trivial topology of the bulk, which exemplifies the bulk-edge correspondence~\cite{Hatsugai_PRL93}~\footnote{
The Chern number computed for the periodic boundary conditions predicts the existence of the chiral edge modes with a boundary, which is known as the bulk-edge correspondence~\cite{Hatsugai_PRL93}
} for two-dimensional systems described by evolutionary game theory; the density profile of player propagates as a wave in the counter-clockwise direction along the edge due to $N_{\mathrm{Ch}}=1$

We note that previous work~\cite{Knebel_RPSchain_PRL20} has studied the dynamics in a one-dimensional system. However, the chiral edge mode protected by the non-trivial topology in the bulk has not been reported so far.

\section{
Summary
}
In this paper, we have discovered the emergence of the chiral edge mode beyond natural science.

Specifically, we have elucidated that the K-RPS hosts a chiral edge mode of the population density which is protected by the non-trivial topology in the bulk.
The emergence of the chiral edge mode is demonstrated by numerically solving the LV equation. 
The dynamics is also intuitively understood by focusing on the ``cyclotron motion" of the single RPS cycle.
Furthermore, we have also elucidated the topological origin of the chiral edge mode by mapping the K-RPS to the fermionic kagome lattice model. 
The former (latter) is described by evolutionary game theory (quantum mechanics).
By making use of the above mapping, we have found that due to the non-zero Chern number of the payoff matrix in the bulk, the chiral edge mode of the population density emerges regardless of the other details of the K-RPS, which exemplifies the bulk-edge correspondence in two-dimensional systems described by evolutionary game theory.
We note that the chiral edge mode induced by bulk topology should also be observed in other systems because its emergence depends only on the non-trivial topology in the bulk~\cite{Haldane_honeycomb_PRL98} 

We finish this paper with two remarks.
We note that a topological band structure in a one-dimensional chain of RPS has been analyzed in Ref.~\onlinecite{Knebel_RPSchain_PRL20}. As well as the discovery of the chiral edge mode, the novelty of this paper is elucidating the mathematical equivalence of the linearized LV equation and the Schr\"odinger equation clarifying the topological origin of the one-way propagating mode observed in the time-evolution.
We also note that the experimental observation of the chiral edge mode is a significant open question to be addressed. Because RPS cycles have been reported for a wide variety of systems, e.g., a system of bacteria~\cite{Kirkup_RPSbac_Nat04}, and human societies~\cite{Semmann_RPSsoc_Nat03,Wang_RPSsoc_SciRep14,Perc_Game_PhysRep17}, we expect the observation of chiral modes in such systems.

\section*{
Acknowledgements
}
This work is supported by JSPS Grant-in-Aid for Scientific Research on Innovative Areas ``Discrete Geometric Analysis for Materials Design": Grants No.~JP20H04627.
This work is also supported by JSPS KAKENHI Grants No.~JP17H06138, No.~JP19K21032, No.~JP20K14371, and No.~JP21K13850.
The authors thank the Supercomputer Center, the Institute for Solid State Physics, University of Tokyo for the use of the facilities.

%


\appendix

\section{
Details of time-evolution for the K-RPS
}
\label{sec: time evol org app}

\begin{figure}[!h]
\begin{minipage}{1\hsize}
\begin{center}
\includegraphics[width=1\hsize,clip]{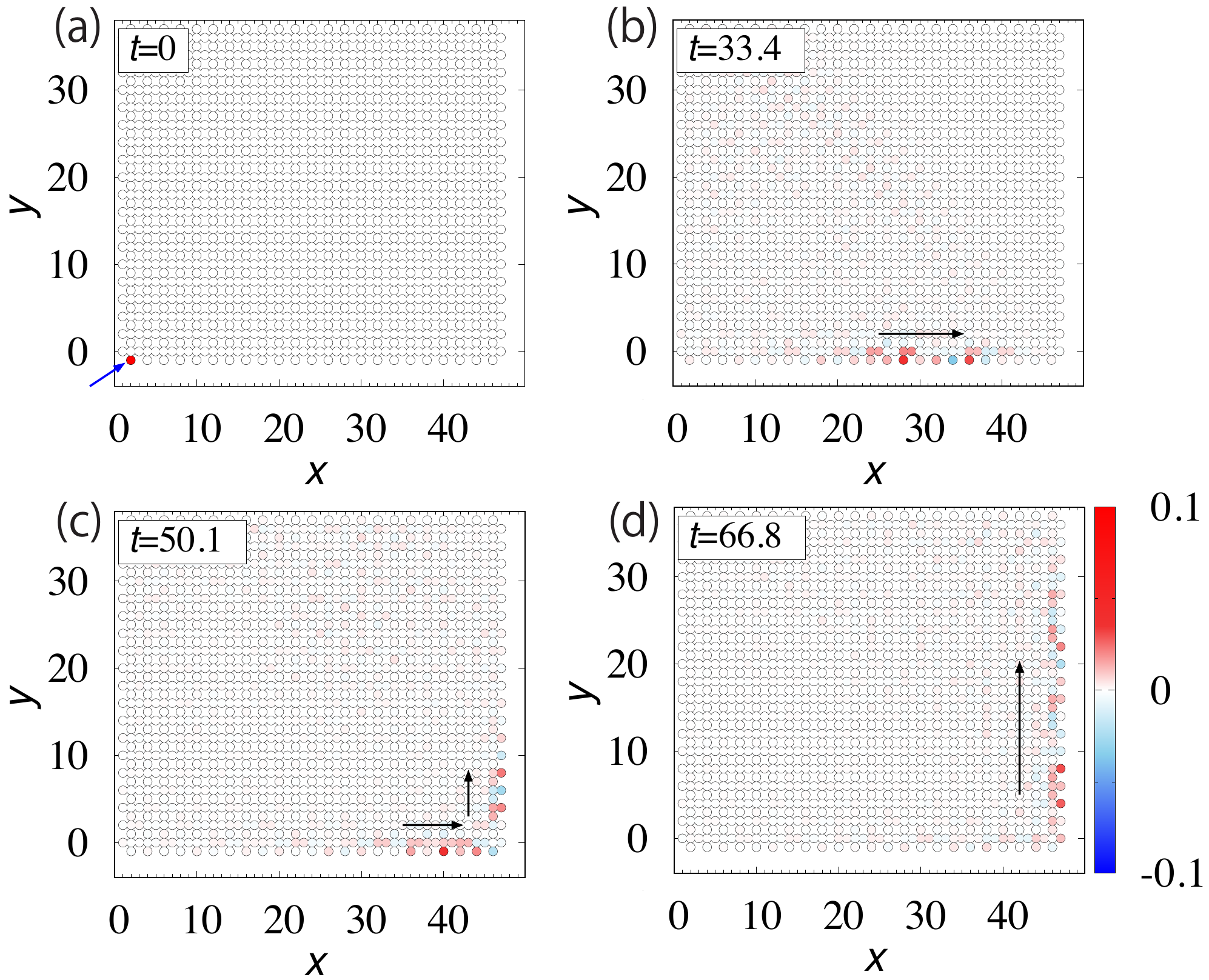}
\end{center}
\end{minipage}
\caption{
(Color Online).
Time-evolution of the population density for the K-RPS.
In these figures, the deviation from $N_{\mathrm{tot}}\bm{c}^{(\mathrm{K})}$ ($\delta \bm{x}=\bm{x}-N_{\mathrm{tot}}\bm{c}^{(\mathrm{K})}$) is plotted; the vector $\bm{x}$ satisfies $\sum_I \bm{x}_I=N_{\mathrm{tot}}+0.1$.
For $t=0$, $\delta x_I$ takes 0.1 at the site denoted by the blue arrow in panel (a); otherwise $\delta x_I$ is zero.
The deviation of the population density propagates along the edge in the counter-clockwise direction as illustrated by the black arrows in panels (b)-(d).
The scale of color plot in panels (a)-(c) is the same as the one in panel (d). 
The data, which are more suited for printing in gray-scale, are provided in Fig.~\ref{fig: Timeevol_K-RPS}.
}
\label{fig: Timeevol_K-RPS org app}
\end{figure}

\textit{
The time-evolution of the population density--
}
In Fig.~\ref{fig: Timeevol_K-RPS}, the sign of deviation is discarded. 
For the complementally information, we plot the deviation $\delta \bm{x}$ ($\delta \bm{x}= \bm{x}-N_{\mathrm{tot}}\bm{c}^{\mathrm{(K)}}$) in Fig.~\ref{fig: Timeevol_K-RPS org app}.

\textit{
Time-evolution for the K-RPS with other conditions--
}
Here, we discuss the dynamics of the K-RPS with three distinct cases; (i) the dynamics with an initial condition significantly deviating from $N_{\mathrm{tot}}\bm{c}^{(\mathrm{K})}$, (ii) the dynamics of the system with a defect on the edge, and (iii) the dynamics of the system where the payoff matrix is replaced as $A\to -A$.

\begin{figure}[!h]
\begin{minipage}{1\hsize}
\begin{center}
\includegraphics[width=1\hsize,clip]{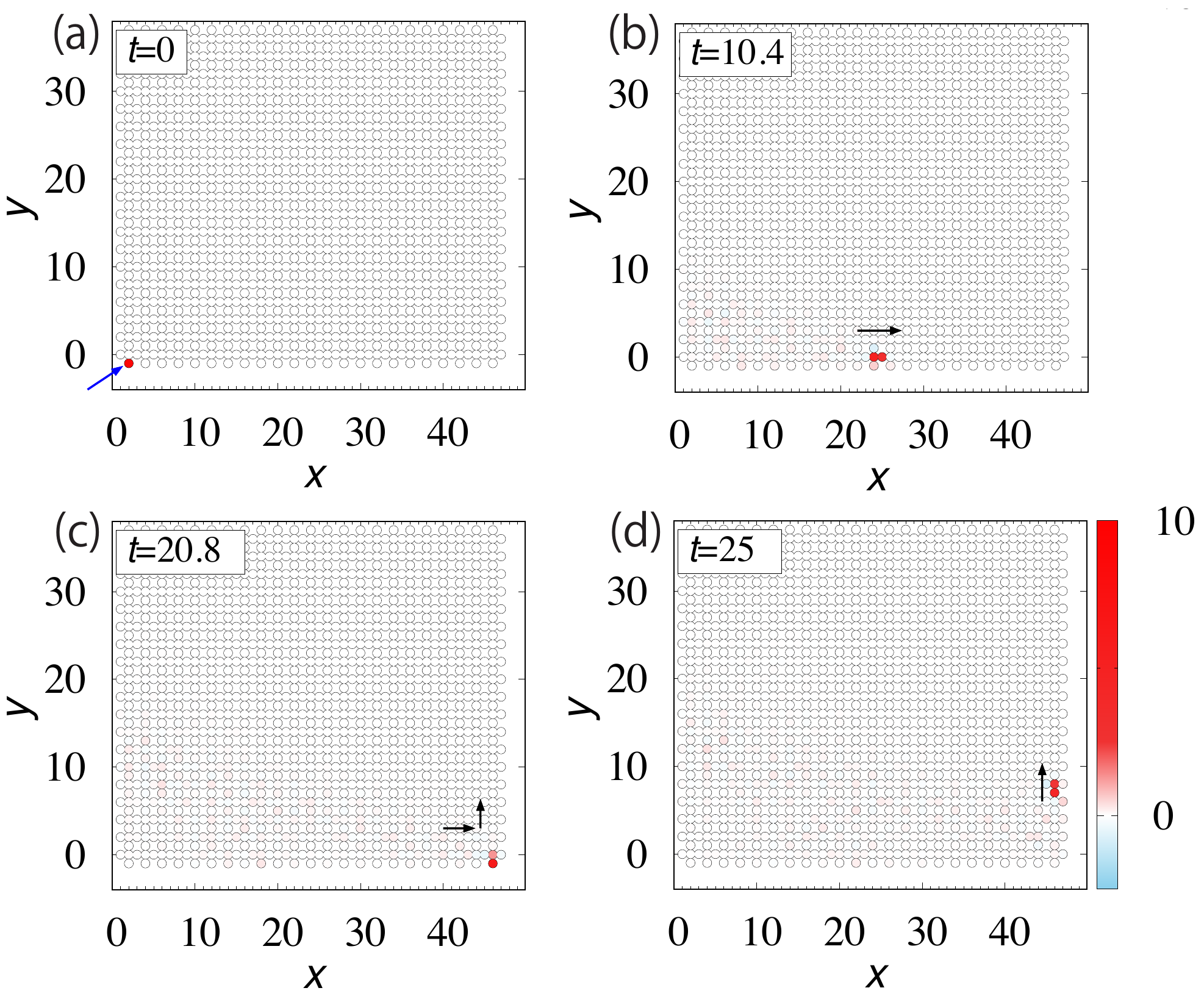}
\end{center}
\end{minipage}
\caption{
(Color Online).
Time-evolution of the population density for the K-RPS. 
In each panel, the deviation from $N_{\mathrm{tot}}\bm{c}^{(\mathrm{K})}$ ($\delta \bm{x}=\bm{x}-N_{\mathrm{tot}}\bm{c}^{(\mathrm{K})}$) is plotted; the vector $\bm{x}$ satisfies $\sum_I \bm{x}_I=N_{\mathrm{tot}}+9$. 
For $t=0$, $\delta x_I$ takes $9$ at the site denoted by the blue arrow in panel (a); otherwise, $\delta x_I$ takes zero. 
The deviation of the population density is positive which propagates along the edge in the counter-clockwise direction as indicated by the black arrows in panels (b)-(d).
The scale of color plot in panels (a)-(c) is the same as the one in panel (d).
}
\label{fig: Timeevol_K-RPS app}
\end{figure}

Firstly, we discuss the dynamics of the K-RPS [see Fig.~1(a)] with an initial condition significantly deviating from $N_{\mathrm{tot}}\bm{c}^{(\mathrm{K})}$.
In Fig.~\ref{fig: Timeevol_K-RPS app}, we can find a one-way propagating mode localized at the edge even when the initial condition is significantly deviates from $N_{\mathrm{tot}}\bm{c}^{(\mathrm{K})}$; the propagation is denoted by the arrows. 

\begin{figure}[!h]
\begin{minipage}{1\hsize}
\begin{center}
\includegraphics[width=1\hsize,clip]{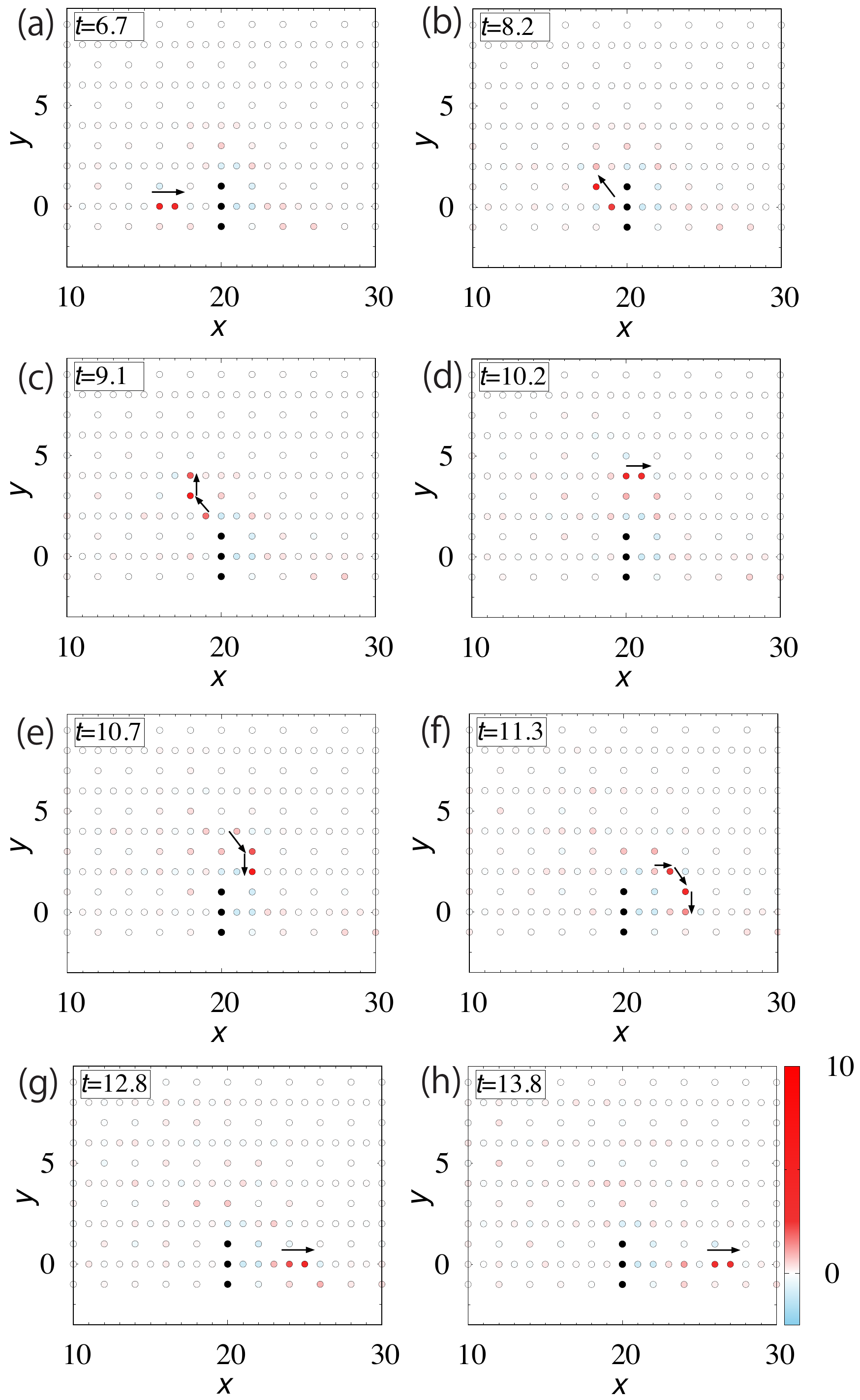}
\end{center}
\end{minipage}
\caption{
(Color Online).
Time-evolution of the population density for the K-RPS with a defect on the edge. 
Sites denoted with black dots are isolated from the other sites, which serves as a defect. Except for the presence of the defect, the set up and the initial condition are 
the same as those of Fig.~\ref{fig: Timeevol_K-RPS app}.
The deviation of the population density is positive whose propagation is denoted by arrows.
The scale of color plot in panels (a)-(g) is the same as the one in panel (e).
}
\label{fig: Timeevol_K-RPS opp2 app}
\end{figure}

Secondly, we demonstrate that even in the presence of a defect on the edge, the chiral edge mode propagates along the edge by detouring around the defect.
Figure~\ref{fig: Timeevol_K-RPS opp2 app} plots the deviation of the population density $\delta x_I$ around the defect. This figure indicates that the population density detours around the defect, which supports the robustness of the chiral edge mode.

\begin{figure}[!h]
\begin{minipage}{1\hsize}
\begin{center}
\includegraphics[width=1\hsize,clip]{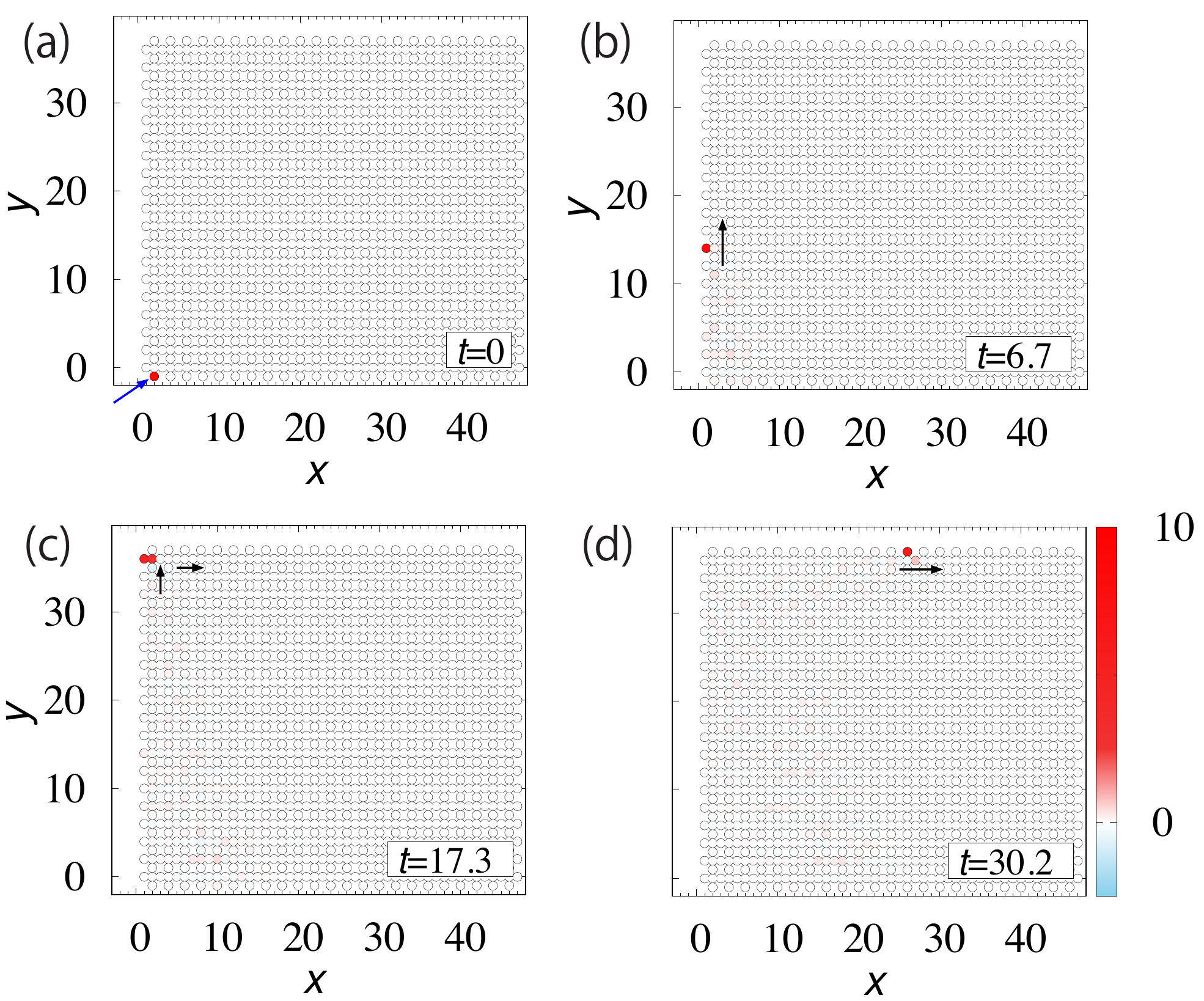}
\end{center}
\end{minipage}
\caption{
(Color Online).
Time-evolution of the population density for the K-RPS. 
Each panel is plotted in a similar way as Fig.~\ref{fig: Timeevol_K-RPS app}.
The deviation of the population density is positive whose propagation in the clockwise direction is denoted by black arrows in panels (b)-(d).
The scale of color plot in panels (a)-(c) is the same as the one in panel (d).
}
\label{fig: Timeevol_K-RPS opp app}
\end{figure}

Thirdly, we discuss the effect of the replacement $A\to-A$ on the dynamics.
As we can see in Fig.~\ref{fig: Timeevol_K-RPS opp app}, the replacement $A\to-A$ flips the chirality of the edge mode, which is consistent with the fact that the replacement $A$ to $-A$ flips the sign of the Chern number.

\section{
Spinless fermions in a kagome lattice with magnetic fluxes
}
\label{sec: fermi app}

The Hamiltonian of spinless fermions in the kagome lattice has been introduced in Ref.~\onlinecite{Ohgushi_ChiralKagome_PRB00}.
However, in order to make this paper self-contained, we briefly describe the model.

Consider spinless fermions in the kagome lattice with magnetic fluxes. The Hamiltonian reads
\begin{eqnarray}
\label{eq: sp fermi app}
H_{\mathrm{fermi}}&=& \sum_{\langle ij\rangle} t_{ij}d^\dagger_{i}d_{j},
\end{eqnarray}
where $d^\dagger_{j}$ ($d_{j}$) creates (annihilates) a spinless fermion at site $j$.
We note that $i$ appearing as subscripts specifies sites.
The hopping integral $t_{ij}$ ($t_{ij}=t^*_{ji}$) describes hopping from site $j$ to site $i$ which takes $t_{ij}=t_0 e^{i\phi}$ when it describes hopping apparel to arrows in Fig.~\ref{fig: K-ele}.
The summation is taken over neighboring sites.

Applying the Fourier transformation, Eq.~(\ref{eq: sp fermi app}) is rewritten as
\begin{widetext}
\begin{subequations}
\label{eq: h(k) fermi app}
\begin{eqnarray}
H_{\mathrm{fermi}} &=& \sum_{\bm{k}} 
\bm{d}^\dagger_{\bm{k}}
h(\bm{k})
\bm{d}_{\bm{k}},
\end{eqnarray}
with
\begin{eqnarray}
h(\bm{k}) &=&
\left(
\begin{array}{ccc}

0 & e^{-i\phi/3} (1+e^{2i\bm{k}\cdot \bm{a}_2}) & e^{i\phi/3} (1+e^{-2i\bm{k}\cdot \bm{a}_1}) \\
h.c. & 0 &   e^{-i\phi/3} (1+e^{2i\bm{k}\cdot \bm{a}_3}) \\
h.c. & h.c.& 0
\end{array}
\right),
\end{eqnarray}
\begin{eqnarray}
\bm{d}_{\bm{k}}&=& 
\left(
\begin{array}{ccc}
d_{\alpha \bm{k}} & d_{\beta \bm{k}} & d_{\gamma \bm{k}}
\end{array}
\right)^T,
\end{eqnarray}
\end{subequations}
\end{widetext}
and $d_{s\bm{k}}=\frac{1}{\sqrt{N_{\mathrm{UC}}} }\sum_{\bm{R}_j} e^{-i\bm{k}\cdot (\bm{R}_{j}) }d_{j}$.
Here, $s$ labels sublattices ($s=\alpha,\beta,\gamma$), and $\bm{R}_j$ denotes a position of a unit cell including site $j$. 
For $\phi=3\pi/2$, Eq.~(\ref{eq: h(k) fermi app}b) is reduced to $iA(\bm{k})$ with $A(\bm{k})$ defined in Eq.~(6b).

We note that Eq.~(\ref{eq: h(k) fermi app}b) is equivalent to Eq.~(5) of Ref.~\onlinecite{Ohgushi_ChiralKagome_PRB00}, which can be seen
by applying the unitary transformation,
\begin{eqnarray}
U&=& 
\left(
\begin{array}{ccc}
1 & 0& 0  \\
0 & e^{-i\bm{k}\cdot \bm{a}_2}&  \\
0 & 0& e^{i\bm{k}\cdot \bm{a}_1}
\end{array}
\right)
\left(
\begin{array}{ccc}
0 &1 &0  \\
0 & 0&1  \\
1 &0 &0
\end{array}
\right).
\end{eqnarray}

We note that the K-RPS can be mapped to the above fermionic system regardless of the number of unit cells. 
Indeed a single-RPS can be mapped to a fermionic three-site system forming a loop. The Hamiltonian reads
\begin{eqnarray}
H_{\mathrm{3sites}}&=& -i
\left(
\begin{array}{ccc}
0  & -1 &  1 \\
1  & 0  & -1 \\
-1 & -1 & 0
\end{array}
\right),
\end{eqnarray}
describing cyclotron motion of a spinless fermion in this three-site system.
The above result elucidate mathematical equivalence between the cyclic motion in the single-RPS and the cyclotron motion in the fermionic three-sites system. 
To be more specific, the linearized LV equation for the single-RPS is mathematically equivalent to the Schr\"odinger equation for the Hamiltonian $H_{3\mathrm{sites}}$.

\section{
Polarization of each eigenstate
}
\label{sec: Kagome_Pol app}

\begin{figure}[!h]
\begin{minipage}{0.7\hsize}
\begin{center}
\includegraphics[width=1\hsize,clip]{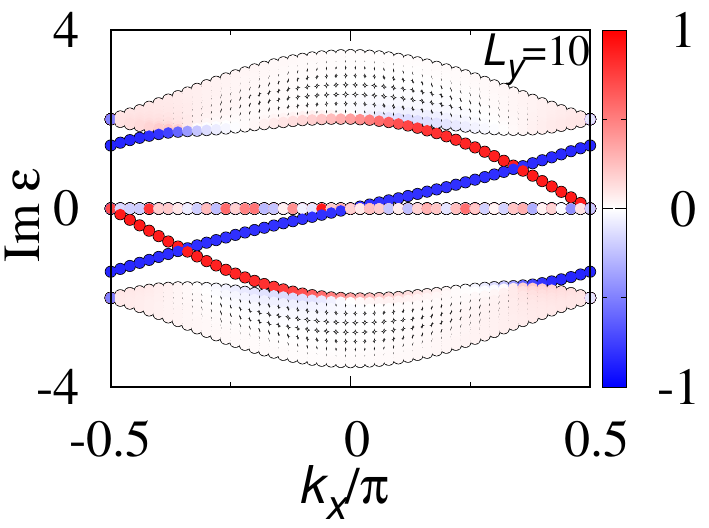}
\end{center}
\end{minipage}
\caption{
Spectrum of the K-RPS under the cylinder geometry. Color of data points represents the polarization $P_n$ defined in Eq.~(\ref{eq: P_n}).
The data, which are more suited for printing in gray-scale, are provided in Fig.~\ref{fig: band structures}(b).
}
\label{fig: band structures app}
\end{figure}

The polarization $P_n$ of each eigenstate is shown in Fig.~\ref{fig: band structures app}.
Noting that $P_n$ takes $P_n=-1$ ($P_n=1$) when the state $\bm{\psi}_{n}(k_x)$ is localized at $J_y=1$ ($J_y=L_y$) [see Eq.~(\ref{eq: P_n})], we can see that the chiral mode denoted by blue-colored (red-colored) dots is localized around $J_y=1$ ($J_y=L_y$) [see Fig.~\ref{fig: band structures app}].

\end{document}